\documentstyle[11pt]{article}
\input{newcoms}
  \advance\oddsidemargin  by -1in
  \advance\evensidemargin by -1in
  \advance\topmargin      by -0.3in
\def\lsim{\mathrel{\rlap{\lower4pt\hbox{\hskip1pt$\sim$}}
    \raise1pt\hbox{$<$}}}     
\def\gsim{\mathrel{\rlap{\lower4pt\hbox{\hskip1pt$\sim$}}
    \raise1pt\hbox{$>$}}}     

\def\beq{\begin{equation}}
\def\endeq{\end{equation}}
\def\arr{\begin{eqnarray}}
\def\endarr{\end{eqnarray}}

\makeindex
\begin{document}
\large
\phantom{.}\hspace{9.4cm} May 28, 1993
\vspace{2cm}
\begin{center}
{\bf \huge Correlation Effects in the Final-State Interaction
for Quasielastic $(e,e'p)$ Scattering \\}
\vspace{1cm}
{\bf N.N.Nikolaev$^{1,2)}$, A.Szczurek$^{1,3)}$,
J.Speth$^{1)}$, J.Wambach$^{1,4)}$, \\
B.G.Zakharov$^{1)}$, V.R.Zoller$^{5)}$ } \medskip\\
{\small \sl
$^{1)}$IKP(Theorie), Forschungszentrum  J\"ulich GmbH.,\\
5170 J\"ulich, Germany \\
$^{2)}$L.D.Landau Institute for Theoretical Physics, \\
GSP-1, 117940, ul.Kosygina 2, V-334 Moscow, Russia\\
$^{3)}$ Institute of Nuclear Physics, PL-31-342 Krakow, Poland\\
$^{4)}$Department of Physics, University of Illinois at
Urbana-Champaign, \\
Urbana, IL 61801, USA\\
$^{5)}$Institute for Theoretical and Experimental Physics, \\
ul.B.Cheremushkinskaya 29, 117129 Moscow, Russia
\vspace{1cm}\\}
{\bf \LARGE A b s t r a c t \bigskip\\}
\end{center}

Color transparency predicts that, in $(e,e'p)$ reactions at
large $Q^2$, the final-state interaction  becomes weaker than the reference
value predicted from the free-nucleon cross
section. This reference value is usually evaluated in the dilute-gas
approximation to Glauber's multiple-scattering theory.
We derive the leading-order correction taking into account two-body
correlations. Large cancellations are found so that the overall
correlation effect is small.

\doublespace

\newpage
\section{ Introduction.}

In the nuclear medium the short-range repulsion between nucleons
leads to correlation effects which influence the intranuclear attenuation
[1]. There is an extensive literature on correlation effects
in elastic hadron-nucleus scattering [2-4] (for a review see [5]).
The recent discussion of color transparency has raised renewed interest in
this phenomenon [6-8]. In order to quantify the strength of the
transparency effect one needs to know the reference value obtained
from in-medium nucleon-nucleon collisions.

For quasielastic $(e,e'p)$ reactions, the subject of
the present communication, the short-range repulsion
leads to the two physical effects:
\begin{itemize}
\item[--]
a local reduction of the nuclear density around the proton
struck by the electron (to be referred to as the {\sl hole} effect).
This reduces the overall attenuation of the struck proton [6,7],
 \item [--]
an enhancement of the final-state interaction of
the struck proton with the correlated spectator nucleons
which leads to an enhancement of the attenuation
(we shall call this the {\sl spectator} effect).

\end{itemize}

The purpose of this paper is twofold: Firstly, we derive the
{\sl spectator} effect which has not been discussed before [9]. We will
find that this effect  largely cancels the {\sl hole} effect.
Secondly, we discuss an additional weakening of the correlation effects
because of the finite (and large) proton-nucleon interaction
radius. This weakening acts differently for the hole and spectator effects.

\section{Correlation effects in the multiple-scattering theory}

Since we are interested in the reference value for nuclear attenuation
we consider correlation effects at moderate $Q^{2}<5\cdot A^{1/3}
\,(GeV/c)^{2}$ where color transparency does not occur [10].
The amplitude for the exclusive process
$e+A\rightarrow e'+ p + (A-1)_{f}$ is given by
\arr
{\cal M}(e+A\rightarrow e'+ p + (A-1)_{f})= ~~~~
{}~~~~~~~~~~~~~~~~~~~~~~~~  \\
\int d\vec{r}_{1}\prod_{j=2}^{A}d\vec{r}_{j}\,\,
\Psi_{(A-1)_{f}}(\vec{r}_{A},...,\vec{r}_{2})^{*}
\Psi_{A}(\vec{r}_{A},...,\vec{r}_{2},\vec{r}_{1})
\exp(i\vec{k}\vec{r}_{1})
S(\vec{r}_{A},......,\vec{r}_{1})G_{em}(Q)
\,\, .                            \nonumber
\label{eq:1}
\endarr
Here $G_{em}(Q)$ is the electromagnetic form factor of the struck proton
at position $\vec r_1$ while $S(\vec{r}_{A},......,\vec{r}_{1})$ denotes the
$S$-matrix element of the intranuclear final-state interaction (FSI)
with the target nucleus debris $\ket{(A-1)_f}$. The reaction kinematics
is shown in Fig.~1. In the semi-exclusive cross section
one sums over all states $\ket{(A-1)_f}$ so that closure can be applied
$
\sum_{f}|(A-1)_{f}\rangle \langle (A-1)_{f}| = 1 ,
$
which gives
\arr
d\sigma_{A} \propto \sum_{f} |{\cal M}|^{2} =
\int d\vec{r}_{1}\,'d\vec{r}_{1}\prod_{j=2}^{A}d\vec{r}_{j}\,\,
\exp\left[i\vec{k}(\vec{r}_{1}-\vec{r}_{1}\,')\right]
G_{em}(Q)^{2} \nonumber\\
\Psi_{A}(\vec{r}_{A},...,\vec{r}_{2},\vec{r}_{1}\,')
S^{\dagger}(\vec{r}_{A},......,\vec{r}_{1}\,')
S(\vec{r}_{A},......,\vec{r}_{1})
\Psi^*_{A}(\vec{r}_{A},...,\vec{r}_{2},\vec{r}_{1}) \, .
\label{eq:2}
\endarr
In the Plane-Wave Impulse Approximation (PWIA) the {\sl r.h.s} of
Eq.~(\ref{eq:2}) gives the single-particle momentum distribution
$dn/d^{3}\vec{k}$. In this case one can fix the initial momentum $\vec k$
of the struck proton from the electron scattering kinematics.
As it is well known, however, the observed $\vec{k}$ distribution will be
different from the single-particle momentum distribution
because of distortions caused by the FSI ($\vec{k}' \neq \vec{k}$,
see Fig.~1). To minimize the distortion effects we consider the $\vec{k}$
integrated cross section.  The resulting transmission coefficient
$Tr_{A}=d\sigma_{A}/Zd\sigma_{N}$ equals
\arr
Tr_{A}=
\int \prod_{j=1}^{A}d\vec{r}_{j}\,\,
|\Psi_{A}(\vec{r}_{A},...,\vec{r}_{2},\vec{r}_{1})|^{2}
S^{\dagger}(\vec{r}_{A},......,\vec{r}_{1})
S(\vec{r}_{A},......,\vec{r}_{1})   \,\, .
\label{eq:3}
\endarr
For vanishing final state interaction, {\it i.e.} $S=1$,
we have $Tr_{A}=1$. To proceed further we use the Glauber expression
for $S^\dagger S$ [1] involving straight-line trajectories of impact
parameter $\vec b$
\beq
S^{\dagger}(\vec{r}_{A},......,\vec{r}_{1})
S(\vec{r}_{A},......,\vec{r}_{1}) =
\prod_{i=2}^{A}[1-\Gamma_{in}(\vec{b}-\vec{c}_{i})]\,\, ,
\label{eq:4}
\endeq
where $\vec{r}_{1}=(\vec{b},z_{1})$, $\vec{r}_{i}=(\vec{c}_{i},z_{i})$ and
\beq
\Gamma_{in}(\vec{b})=
\Gamma^*(\vec{b})+
\Gamma(\vec{b})-
\Gamma^*(\vec{b})
\Gamma(\vec{b}) \approx
{\sigma_{in}(pN)\over 2\pi B_{in}}
\exp\left(-{b^{2}\over 2B_{in}}\right) .
\label{eq:5}
\endeq
Here $\Gamma(\vec{b})$ is the profile function for proton-nucleon
elastic scattering. For all practical purposes $B_{in}$ is identical to
the diffraction slope $B_{pN}$ for elastic proton-nucleon scattering (the
two slopes differ by terms $\sim \sigma_{el}(pN)/\sigma_{tot}(pN)
\sim 0.2$).

In the dilute-gas approximation the transparency factor takes the
simple form [10]
\arr
Tr_{A}^{(0)}= {1\over A}\int d^{2}\vec{b} dz_{1}n_{A}(\vec b,z_{1})
\left[1-{1\over A}\sigma_{in}(pN)t(\vec b,z_{1})\right]^{A-1}\nonumber\\
\approx {1\over A}\int d^{2}\vec{b} dz_{1}n_{A}(\vec b,z_{1})
\exp[-\sigma_{in}(pN)t(\vec b,z_1)] ,
\label{eq:6}
\endarr
where $t(\vec b,z_{1})$ is the optical thickness:
\beq
t(\vec b,z_{1})=\int_{z_{1}}^{\infty} dz n_{A}(\vec b,z)
\label{eq:7}
\endeq
and $n_A(\vec r)$ the mass density of the nucleus.
The $z_{1}$-integration can be performed explicitly and
\arr
Tr_{A}^{(0)}
= {1\over A\sigma_{in}(pN)}\int d^{2}\vec{b} \left\{1-
\left[1-{1\over A}\sigma_{in}(pN)T(\vec b)\right]^{A} \right\}
\nonumber\\
\approx {1\over A\sigma_{in}(pN)}\int d^{2}\vec{b} \left\{1-
\exp\left[- \sigma_{in}(pN)T(\vec b)\right] \right\} \,\, ,
\label{eq:8}
\endarr
where $T(\vec b)=t(\vec b,-\infty)$.
For simplicity we shall use the exponentiated form  in the
following.

To discuss higher-order corrections, consider the $\nu$-fold
{\it rescattering} contribution to $Tr_A$ in Eq.~(3):
\beq
Tr_{A}^{(\nu)}=
{(-1)^{\nu}(A-1)! \over (A-1-\nu)!\nu!}
\int d^{2}\vec{b}dz_{1}\int_{z_{i}>z_{1}}\prod_{i=2}^{\nu+1}
d^{3}\vec{r}_{i}
\prod_{i=2}^{\nu+1}\Gamma_{in}(\vec{b}-\vec{c}_{i})\,
n^{(\nu+1)}(\vec{r}_{\nu+1},....,\vec{r}_{1})
\label{eq:9}
\endeq
which involves the $(\nu+1)$-body density
$n^{(\nu+1)}(\vec{r}^{\nu+1},...,\vec{r}_{1})$. We define it such
that it satisfies the sum rules
\beq
\int d^{3}\vec{r}_{\nu}n^{(\nu)}(\vec{r}_{\nu},....,\vec{r}_{1})=
n^{(\nu-1)}(\vec{r}_{\nu-1},....,\vec{r}_{1}) \, .
\label{eq:10}
\endeq
(Here we have used the Foldy-Walecka convention [11].) It should be
noted that the multiplicity $\nu$ of rescatterings is small compared
to the mass number of the nucleus. An estimate can be obtained from the
proton-nucleon inelastic cross section. With $\sigma_{in}(pN)\approx 32 mb$
one gets
\beq
\nu \lsim \sigma_{in}(pN)n_{A}(0)R_{A}
\approx  0.7\,A^{1/3} \ll A   \,.
\label{eq:11}
\endeq
Therefore the principle correction to the dilute-gas approximation comes
from two-body correlations which involve the two-body density
\beq
n^{(2)}(\vec{r}_{2},\vec{r}_{1})=
n^{(1)}(\vec{r}_{2})n^{(1)}(\vec{r}_{1})
\left[1-C(\vec{r}_{2},\vec{r}_{1})\right]   \, ,
\label{eq:12}
\eeq
where $n^{(1)}(\vec{r})\equiv n_{A}(\vec r)/A$ .
By virtue of the normalization condition (\ref{eq:10})
the so-defined correlation function $C$ satisfies the sum rule
[11]
\beq
\int d^{3}\vec{r}_{1}
n^{(1)}(\vec{r}_{1})
C(\vec{r}_{2},\vec{r}_{1})=0  \,.
\label{eq:13}
\endeq
Including two-body correlations we can decompose the $\nu$-body density
as
\bea
n^{(\nu)}(\vec{r}_{\nu},....,\vec{r}_{1})=
\prod_{j=1}^{\nu} n^{(1)}(\vec{r}_{j})
\prod_{i>k}[1-C(\vec{r}_{i},\vec{r}_{k})]\nonumber\\
 \approx
\left[1-\sum_{i>1}C(\vec{r}_{i},\vec{r}_{1})-
\sum_{i>k>1}C(\vec{r}_{i},\vec{r}_{k})\right]
\prod_{j=1}^{\nu} n^{(1)}(\vec{r}_{j})  \, .
\label{eq:14}
\eea
After the $\nu$-body density (14) is substituted into Eq.(\ref{eq:9}),
the zeroth order terms reproduce the transmission coefficient
$Tr_A^{(0)}$ in Eq.(\ref{eq:8}). The effect of the correlation hole
(the {\sl hole effect})
described by terms $\propto \sum_{i>1}C(\vec{r}_{i},\vec{r}_{1})$ leads
to the correction to the nuclear transparency
\bea
\Delta Tr_{A}^{(h)} = {1 \over A} \sum_{i>1}
\int d^{2}\vec{b}dz_{1} n_{A}(\vec b,z_{1})
\exp\left[-\sigma_{in}(pN)t(\vec b,z_{1})\right] \nonumber\\
\int_{z_{i}>z_{1}} d^{3}\vec{r}_{i}
n^{(1)}(\vec{r}_{i})  \Gamma_{in}(\vec{b}-\vec{c}_{i})
C(\vec{r}_{1},\vec{r}_{i}) .
\label{eq:14a}
\eea
The terms $\propto \sum_{i>k>1}C(\vec{r}_{i},\vec{r}_{k})$ describe
correlations between two spectator nucleons (the {\sl spectator effect})
and give rise to the correction
\bea
\Delta Tr_{A}^{(s)} = - {1 \over A} \sum_{k>i>1}
\int d^{2}\vec{b}dz_{1} n_{A}(\vec b,z_{1})
\exp\left[-\sigma_{in}(pN)t(\vec b,z_{1})\right] \nonumber\\
\int_{z_{i},z_{k}>z_{1}} d^{3}\vec{r}_{i} d^{3}\vec{r}_{k}
n^{(1)}(\vec{r}_{i}) n^{(1)}(\vec{r}_{k})
\Gamma_{in}(\vec{b}-\vec{c}_{i}) \Gamma_{in}(\vec{b}-\vec{c}_{k})
C(\vec{r}_{i},\vec{r}_{k}) .
\label{eq:14b}
\eea

\section{The spectator effect}

There are $(A-1)(A-2) \approx A^2$ identical terms in Eq.~(\ref{eq:14b})
, thus it is enough to calculate
\beq
S_{23} = \int_{z_{2},z_{3}>z_{1}} d^{3}\vec{r}_{2} d^{3}\vec{r}_{3}
n^{(1)}(\vec{r}_{2}) n^{(1)}(\vec{r}_{3})
\Gamma_{in}(\vec{b}-\vec{c}_{2}) \Gamma_{in}(\vec{b}-\vec{c}_{3})
C(\vec{r}_{2},\vec{r}_{3}) .
\label{eq:16}
\endeq
To evaluate this integral approximately
we notice that the proton-nucleon interaction radius is much smaller
than the radius of the nucleus ($B_{in} \ll R_{A}^{2}$).
Therefore, according to (5),
$\Gamma_{in}(\vec{b}-\vec{c}_{i})$ are strongly peaked
around $\vec{c}_{i}=\vec{b}$ which implies that
$n^{(1)}(\vec{r}_{i}) = n^{(1)}(\vec c_{i},z_{i})
\approx n^{(1)}(\vec b,z_{i})$.
Secondly, the correlation radius $r_{c}$ is also much
smaller than $R_A$ and therefore
$n^{(1)}(\vec b,z_{3})\approx n^{(1)}(\vec b,z_{2})$. For the same reason
one can take
$C(\vec{r}_{1},\vec{r}_{2})=C(\vec{r}_{1}-\vec{r}_{2})$.
Then, using the explicit form of the profile function
$\Gamma_{in}(\vec b)$ in Eq.~(\ref{eq:5}), we have (here
$(\vec{c},z)=\vec{r}_{2}-\vec{r}_{3}$)
\arr
S_{23} \approx \sigma_{in}(pN)^{2}
\int_{z_1} dz_{2}
n^{(1)}(\vec b,z_{2})^{2}\int dz{1\over 4\pi B_{in}}
\int d^{2}\vec{c}\,\exp\left(-{\vec{c}^{2}\over 4B_{in}}\right)
C(\vec{c},z)
\nonumber\\
\approx {1\over A^{2}}\sigma_{in}(pN)^{2}
\int_{z_1} dz_{2} n_{A}(\vec{b},z_{2})^{2}
{r_{c}^{2} \over r_{c}^{2}+2B_{in}}
2\int_{0}^{\infty} dr\,C(r) \, \, .~~~~~~~
\label{eq:17}
\endarr
In deriving the approximate expression in the last line we have
used a Gaussian form
$C(\vec{r}_{2},\vec{r}_{3})\approx
\exp[-(\vec{r}_{2}-\vec{r}_{3})^{2}/2r_{c}^{2}]$
which neglects small corrections from the long-range attractive part
of the correlation function (see the sum rule (\ref{eq:13})).
Their inclusion would slightly lower the correlation effect.
Notice the weakening of the correlation effect from the finite $pN$
interaction radius in Eq.~(17). A similar weakening has been derived
in [3,4] for the elastic hadron-nucleus scattering.
The final expression for the correction to $Tr_A$ from the
repulsion of the spectator nucleons is then given by
\beq
\Delta Tr_{A}^{(s)}\approx
-{1 \over A}
\sigma_{in}(pN)^{2}
l_{cor}
\int d^{2}\vec{b} dz_{1} n_{A}(\vec{b},z_{1})
\exp\left[-\sigma_{in}(pN)t(\vec b,z_1)\right]
\int_{z_{1}}^{\infty} dz_{2}n_{A}(\vec{b},z_{2})^{2}\, ,
\label{eq:18}
\endeq
where $l_{cor}$ is the effective correlation length
\beq
l_{cor}={r_{c}^{2} \over r_{c}^{2}+2B_{in}}\int_{0}^{\infty} drC(r)
\approx \sqrt{{\pi\over 2}}r_{c}{r_{c}^{2} \over r_{c}^{2}+2B_{in}} .
\label{eq:19}
\endeq
When combined with the dilute-gas result
the spectator effect corresponds to a multiplicative factor
\beq
1-
\sigma_{in}(pN)^{2} l_{cor}
\int_{z_{1}}^{\infty}dz n_{A}(\vec b,z)^{2}
\label{eq:20}
\endeq
in the integrand in Eq.~(\ref{eq:6}).
Above we have considered one correlated spectator pair. The extension
to the higher-order terms $\propto C(\vec{r}_{i},\vec{r}_{j})
C(\vec{r}_{k},\vec{r}_{l})$ is straightforward.
After they are included, the correction factor (\ref{eq:20})
exponentiates and one obtains a compact result: the optical thickness
function $t(\vec b,z_1)$ in the dilute-gas expression
(\ref{eq:6}) is simply  replaced by
\beq
t(\vec b,z_1)=\int_{z_{1}}^{\infty}dz n_{A}(\vec b,z)
\left[1+l_{cor}\sigma_{in}(pN)n_{A}(\vec b,z)\right] \,\, .
\label{eq:21}
\endeq
Evidently, the spectator effect enhances the attenuation and reduces
nuclear transparency.

\section{The hole effect}

The contribution of $C(\vec r_1,\vec r_i)$ to $Tr_A$ is evaluated in
analogy to (\ref{eq:16}) and (\ref{eq:17}) and yields
(here $(\vec{c},z)=\vec{r}_{1}-\vec{r}_{2}$)
\arr
\Delta Tr_{A}^{(h)} \approx
{1 \over A}
\int d^{2}\vec{b}dz_{1}
n_{A}(\vec b,z_{1})^{2}
\exp\left[-\sigma_{in}(pN)t(\vec b,z_{1})\right]
\int_{0}^{\infty} dz
\int d^{2}\vec{c}\,
\Gamma_{in}(\vec{c})
C(\vec{c},z) ~~~~~~~~~  \nonumber\\
\approx
{1\over A}
\sigma_{in}(pN)
{r_{c}^{2}\over r_{c}^{2}+B_{in}  }\int dr C(r)
\int d^{2}\vec{b}
dz_{1}n_{A}(\vec{b},z_{1})^{2}
\exp\left[-\sigma_{in}(pN)t(\vec b,z_{1})\right]\, .
{}~~~~~~~~~~
\label{eq:22}
\endarr
Also the hole effect is weakened by the finite radius of the $pN$
interaction although less than for the spectator effect.

When combined with the dilute-gas formula (\ref{eq:6}), the hole
correction (\ref{eq:22}) enhances the integrand (\ref{eq:6})
 by a factor
\beq
1+
\int_{z_{1}}^{\infty}dz n_{A}(\vec b,z)
\int d^{2}\vec{b} \Gamma_{in}(\vec{b})C(\vec{b},z-z_{1}) \approx
1+\sigma_{in}(pN)l_{cor}{r_{c}^{2}+2B_{in} \over
r_{c}^{2}+B_{in}} n_{A}(b,z_{1})
\,\, .
\label{eq:23}
\endeq
Unlike the spectator correction, however, the hole correction
does not exponentiate (here we disagree with ref.~[6] in principle).
But as long as the correction factor (\ref{eq:23}) does not differ much
from unity, it may as well be written in exponential form. In
this case the combined effect of the spectators and the hole on $Tr_A$
that is $Tr_{A}=Tr_{A}^{0}+\Delta Tr_{A}^{(s)}+
\Delta Tr_{A}^{(h)}$ can be represented entirely by a modification of the
optical thickness function $t(\vec b,z_1)$ entering  Eq.~(\ref{eq:6}).
One has
\beq
t(\vec b,z_1)=\int_{z_{1}}^{\infty}dz n_{A}(\vec b,z)
\left[1+\sigma_{in}(pN)l_{cor}n_{A}(\vec b,z)
-{1 \over \sigma_{in}(pN) }
\int d^{2}\vec{b} \Gamma_{in}(\vec{b})C(\vec{b},z-z_{1})
\right] \,\, ,
\label{eq:24}
\endeq
which clearly shows the cancellation between the spectator and hole
effects.

\section{Numerical estimates}

The $pN$ scattering parameters change little above few GeV energy.
In the following calculations we have used
$\sigma_{in}(pN)= 32 mb$, $B_{in}=0.5$ fm$^{2}$ [12] and
$r_{c}=0.5$ fm [13,7].
In the upper panel of Fig.~2 we present numerical results for
$\Delta Tr_A^{(h)}/Tr_A^{(0)}$ and
$\Delta Tr_A^{(s)}/Tr_A^{(0)}$ as well as the net correction (sum
of the both).
The finite range of high-energy $pN$ interaction leads to
a rather strong reduction of the correlation effect:
$r_{c}^{2}/(r_{c}^{2}+2B_{in}) \approx 1/5 $ in the
spectator effect (Eq.~(\ref{eq:17})),
and $r_{c}^{2}/(r_{c}^{2}+B_{in}) \approx 1/3$
in the hole effect (Eq.~(\ref{eq:22})).
We find (see the lower panel of Fig.2)
very small overall correlation effects, so that the dilute-gas
approximation gives reliable estimates for the nuclear
attenuation. Benhar et al. [6] did not consider the spectator effect
and quote a numerically larger hole effect (apparently,
their predictions are for $B_{in}\ll r_{c}^{2}$, in which case
we find agreement with ref.~[6]).

The above numerical estimates apply for $Q^{2}
\lsim 5\,A^{1/3}\,(GeV/c)^{2}$ which is below the onset of  color
transparency effects [10]. Extensions to the regime of color
transparency go beyond the scope of this paper and will be discussed
separately.

\section{Conclusions}

We have investigated the role of two-body correlations in nuclear
transparency for the quasielastic $(e,e'p)$ reaction.
We find large cancellations of the spectator and hole effects.
The correlation effects are further reduced
by the finite range of the $pN$ interaction.
These conclusions are also relevant for $(p,2p)$ reactions.
We conclude that the lowest-order Glauber model estimates of nuclear
attenuation are rather accurate. After this work has been completed,
we became aware of a paper by Kohama, Yazaki and Seki [14] on the
correlation effects in $(e,e'p)$ scattering on the $^4$He. They also
find a small correlation effect in agreement with the discussion
[15] for elastic proton-$^4$He scattering.

This work was supported in part by the Polish KBN grant 2 2409 9102.

\pagebreak

{\bf Figure captions}:
\begin{itemize}

\item[ Fig.1 - ]
        The kinematics of $eA \rightarrow e'pA_{f}$ reaction including
        final state interaction.

\item[ Fig.2 - ]
        Correlation effect in the nuclear transmission coefficient
        $Tr_A$ as a function
        of mass number $A$. The upper part displays the relative
        corrections $\Delta Tr^{(h)}_A/Tr_A^{(0)}$ (hole effect) and
        $\Delta Tr^{(s)}_A/Tr_A^{(0)}$ (spectator effect), as well as
        their sum (net effect). In the lower part is shown the
        transmission factor $Tr_A$ without correlation effects (dashed
        line) and with two-body correlations (full line).

\end{itemize}
\end{document}